\documentclass[twocolumn]{aastex63}

\submitjournal{ApJ}

\shorttitle{Can we trust MHD jump conditions for collisionless shocks?}
\shortauthors{Bret}

\begin{document}

\title{Can we trust MHD jump conditions for collisionless shocks?}

\correspondingauthor{Antoine Bret}
\email{antoineclaude.bret@uclm.es}

\author[0000-0003-2030-0046]{Antoine Bret}
\affiliation{ETSI Industriales, Universidad de Castilla-La Mancha, 13071 Ciudad Real, Spain}
\affiliation{Instituto de Investigaciones Energ\'{e}ticas y Aplicaciones Industriales,\\Campus Universitario de Ciudad Real,  13071 Ciudad Real, Spain}

\begin{abstract}
When applied to compute the density jump of a shock, the standard magnetohydrodynamic (MHD) formalism assumes, 1) that all the upstream material passes downstream, together with the momentum and energy it carries, and 2) that pressures are isotropic. In a collisionless shock, shock accelerated particles going back and forth around the front can invalid the first assumption. In addition, an external magnetic field can sustain stable pressure anisotropies, invaliding the second assumption. It is therefore unclear whether  the density jump of a collisionless shock fulfils the MHD  jump or not.

Here we try to clarify this issue. A literature review is conducted on 68 articles dealing with Particle-In-Cell simulations of collisionless shocks. We analyze the factors triggering departure from the MHD density jump and quantify their influence on $\Delta_{RH}$, the relative departure from the Rankine-Hugoniot jump. For small departures we  propose $\Delta_{RH} = + \mathcal{O}(10^{-1-3.7\kappa})t^\kappa - \sigma \mathcal{O}(1)$ where $t$ is the timescale of the simulation,  $\sigma$ the magnetization parameter and   $\kappa$ a constant of order unity.  The first term stems from the energy leakage into accelerated particle. The second term stems from the downstream anisotropy triggered by the field (assuming an isotropic upstream). This relation allows to assess to which extent  a collisionless shock  fulfils the RH density jump.

In the strong field limit and for parallel shocks, the departure caused by the field saturates at a finite, negative, value. For perpendicular shocks, the departure goes to zero at small and high $\sigma$'s so that we find here a departure window.   The results obtained have to be checked against full 3D simulations.
\end{abstract}


\section{Introduction}

Since their discovery during the 19th century \citep{johnson1998classic,Salas2007}, shockwaves have been the object of innumerable investigations. The fluid equations first used to describe them operate under the assumption that the mean-free-path of the particles  is much smaller than any other dimension of the system under scrutiny. With such a prominent role given to binary collision to randomize the flow at the microscopic level, it is reasonable to assume 1) that the pressure is isotropic in both the upstream and the downstream and 2) that all the matter upstream goes downstream, together with the energy and the momentum it carries\footnote{Radiative shocks \citep{Zeldovich,mihalas1999} are excluded from the discussion.}. The second assumption allows to apply the conservation laws between the upstream and the downstream, while the first assumption allows to write these laws using fluid mechanics or magnetohydrodynamics (MHD) equations. From there, one derives the jump conditions for the density, pressure, magnetic field, etc. \citep{fitzpatrick2014plasma,keppens2019}.

Contrary to fluid shockwaves, were dissipation at the shock front is provided by binary collisions, collisionless shockwaves are mediated by collective plasma effects on length scales much shorter than the mean-free-path \citep{Sagdeev66,tidman1971shock,balogh2013}. A good example is the earth bow-shock in the solar wind, where the shock front is about 100 km thick while the mean-free-path at the same location is of the order of the sun-earth distance \citep{PRLBow1,PRLBow2}.

In the absence of binary collisions to isotropize the flow, to which extent can we assume isotropic pressures? Also, given the mean-free-path is much larger than the shock front, to which extent can we assume all the matter upstream goes downstream, together with the momentum and energy it carries? Indeed, it turns out that these two assumptions  are  far from obvious in a collisionless environment. As a consequence, it is not obvious either that the fluid or MHD jump conditions derived for a collisional fluid, are still valid.

Note that we hereafter refer to MHD jump conditions derived considering \emph{isotropic} pressures. Several authors adapted them to the case of anisotropic pressures, considering the anisotropy degree as a free parameter \citep{Karimabadi95,Erkaev2000,Vogl2001,Gerbig2011}. Yet, the goal of the present paper is to compare jump conditions (mainly the density jump) of collisionless shocks with the simple and well known ``isotropic MHD'' jump conditions, also frequently referred to as ``Rankine-Hugoniot'' (RH) jump conditions, even though William Rankine and Pierre Hugoniot derived these relations for a neutral fluid. In the sequel, we shall use interchangeably ``isotropic MHD'', ``MHD'' or ``RH''.

Two processes have been identified  that can trigger a non-RH density jump,

$\bullet$ An external magnetic field $\mathbf{B}_0$ can sustain stable anisotropies, breaking the isotropy assumption of MHD. Its strength is characterized by the $\sigma$ parameter,
  \begin{equation}\label{eq:sigma}
  \sigma = \frac{B_0^2/4\pi}{(\gamma_1-1)n_1 (\sum_im_i) c^2},
  \end{equation}
  where $n_1$ and $\gamma_1$ are respectively the upstream density and Lorentz factor (measured in the downstream frame). The  $m_i$'s are the masses of the species composing the plasma.

$\bullet$ As they accelerate particles, collisionless shocks generate a population which goes back and forth around the front, breaking the ``everything upstream goes downstream'' assumption. As we shall see in Section \ref{sec:accel}, the process can be characterized by the parameter,
     \begin{equation}\label{eq:alpha}
     \alpha = \frac{F_E}{\frac{1}{2}n_1v_1^3},
     \end{equation}
      where $F_E$ is the energy fluxes escaping the Rankine-Hugoniot budget  and $v_1$ the upstream velocity.

Particle-in-cell (PIC) simulations are undoubtedly the tool \emph{par excellence} to study non-linear collisionless phenomena like collisionless shocks. Because they operate from first principles at the microscopic level, they are inherently kinetic. We therefore present in Section \ref{sec:litt} a literature review of PIC simulations of collisionless shocks, magnetized or not, in pair or electron/ion plasmas. The observations gathered will then feed Section \ref{sec:theo} where departures from MHD density jump are modelled.

Most of the simulations found in literature use the ``reflecting wall'' technique to produce a shock. There, a semi-infinite plasma is sent against a reflecting wall where it bounces back to interact with itself. The present work focuses on this technique. In this reflecting scheme, the simulations are therefore performed in the downstream frame of the formed shock. By design, such a scheme can only simulate shocks formed by the encounter of 2 identical plasmas.

Noteworthily, the less represented ``injection method'' allows to study shocks produced by the collision of any 2 kinds of plasmas (different compositions and/or different densities). Shocks arising from the interaction of a jet with a standing plasma can be studied with this scheme. For example \cite{NishikawaApJL2009} could study the interaction of a diluted relativistic pair jet with a unmagnetized pair plasma. While many ``reflecting wall papers'' studied shocks in pair plasmas (see Sections \ref{sec:littnoB0} and \ref{sec:pairsB0}), a density ratio different from unity (\cite{NishikawaApJL2009} has 0.676) is only achievable with the injection method. Still with the injection method, \cite{Ardaneh2016} studied the interaction of an electron jet with an unmagnetized electron/ion plasma, and commented on the differences between the reflected and injected schemes. Magnetized systems have also been explored, with \cite{DieckmannAA2019}, for example, considering a pair jet colliding with an electron-proton plasma over a guiding magnetic field.

As is appears, the injection method truly allows for an extensive exploration of the possible shocks. The reflected wall scheme restricts the dimension of the parameters phase space, and to date counts with more studies, which is why we here focus on it. Yet, it would be interesting to extend the current analysis to the injection scheme.

Defining now the density ratio between the shock upstream (subscript ``1'') and downstream (subscript ``2'') like,
\begin{equation}\label{eq:r}
r=\frac{n_2}{n_1},
\end{equation}
we shall model $\Delta_{RH}$, the relative departure from the RH jump $r_{RH}$, defined by,
\begin{equation}\label{eq:delta0}
\Delta_{RH}(\sigma, \alpha) \equiv \frac{r - r_{RH}}{r_{RH}}.
\end{equation}

Beyond the elaboration of a full theory of the density jump accounting for the effects listed above, our present goal is mainly to determine when the RH density jump does apply to collisionless shocks.

\section{Literature review}\label{sec:litt}
We conducted a literature review of PIC simulations of collisionless shocks. We selected 68 articles where 1) a shock structure was clearly obtained, with a downstream significantly longer than the overshoot region, if any, right behind the front, and 2) the density jump can be related to its MHD counterpart with a reasonable accuracy, whether explicitly or implicitly. The medium where the shock propagates is homogenous (see \cite{2019ApJ...886...54T} for  an inhomogeneous case). Save a few exceptions like \citet{2009ApJ...698.1523S,2012PPCF...54l5004S,2018MNRAS.477.5238P,2018ApJ...858...95G}, the density jump was not explicitly compared to its MHD counterpart, for such was not the main goal of the article. It is then possible that a few percent discrepancy between the 2 went unnoticed for some articles.

\subsection{Un-magnetized shocks}\label{sec:littnoB0}
All the articles examined but \cite{2009ApJ...693L.127K, 2012PPCF...54l5004S} pertaining to the un-magnetized regime, relativistic or not, display a shock in agreement with the MHD requirements.

\emph{For shocks in pair plasmas} \citep{2005AIPC..801..345S, 2007ApJ...668..974K, 2008ApJ...674..378C, 2008ApJ...682L...5S, 2009ApJ...693L.127K, 2009ApJ...707L..92S, 2013PhPl...20d2102B, 2014PhPl...21g2301B, 2016PhPl...23f2111D, 2017JPlPh..83a9004D, 2017PhPl...24d2113L, 2018MNRAS.473..198D, 2019PhRvE.100a3205P, 2019PhRvE.100c3209L, 2019PhRvE.100c3210L, 2019PhRvL.123c5101L, 2020arXiv200201141V}, the longest simulation \citep{2009ApJ...693L.127K} was ran up to $11925\omega_p^{-1}$, where $\omega_p$ is the electronic plasma frequency.

In \cite{2012PPCF...54l5004S} the authors carefully measured the departure from MHD, as the very goal of the paper was to ``assess the impact of non-thermally shock-accelerated particles on the MHD jump conditions of relativistic shocks''. Pushing the simulation up to $2395\omega_p^{-1}$, a $+7\%$ departure for the density jump was found.

Although \cite{2009ApJ...693L.127K} did not precisely measured the density jump,  Figure 1 of their article shows a MHD jump at $2250\omega_p^{-1}$ and a slight departure ($+3.5\%$) at $11925\omega_p^{-1}$ due to energy leakage in accelerated particles (see Section \ref{sec:accel}).

\emph{For shocks in electron/ion plasmas} \citep{2008ApJ...673L..39S, 2008ApJ...681L..93K, 2009ApJ...695L.189M, 2010PPCF...52b5001D, 2012ApJ...759...73N, 2012PhRvL.108w5004F, 2014NatSR...4E3934S, 2014PhRvL.113j5002S, 2015PhPl...22h2107R, 2015ApJ...803L..29S, 2017PhPl...24d1409R, 2018PhPl...25a2118N, 2020PPCF...62b5022M}, the longest simulation time was $4111 \omega_{pi}^{-1}$ \citep{2012ApJ...759...73N}. No significant departure from the MHD jump was detected in any article.

The case of pair plasmas suggests that departure from MHD due to accelerated particles requires running the simulation for several thousands of electronic plasma frequencies to be perceptible. In electron/ion plasmas, this translate to running the simulation for several thousands of ionic plasma frequencies. The longest run examined in this respect was $4111\omega_{pi}^{-1}$ \citep{2012ApJ...759...73N}, where $\omega_{pi}$ is the ionic plasma frequency. Yet the density jump is not measured accurately enough\footnote{See footnote 13 for more on particle acceleration in \cite{2012ApJ...759...73N}.}. Pushing simulations beyond this time scale for non-MHD effects to become clear, requires so far Hybrid simulations discussed in Section \ref{sec:hybrid}.

An important feature observed is related to accelerated particles. Their effect on the shock is not steady. As specified in \cite{2009ApJ...693L.127K},  ``simulations do not reach a steady state; rather, an increasing fraction of shock energy is transferred to energetic particles and magnetic fields throughout the simulation time domain''. We shall comment further on this point in Section \ref{sec:accel}.

  \begin{figure}
\begin{center}
 \includegraphics[width=0.45\textwidth]{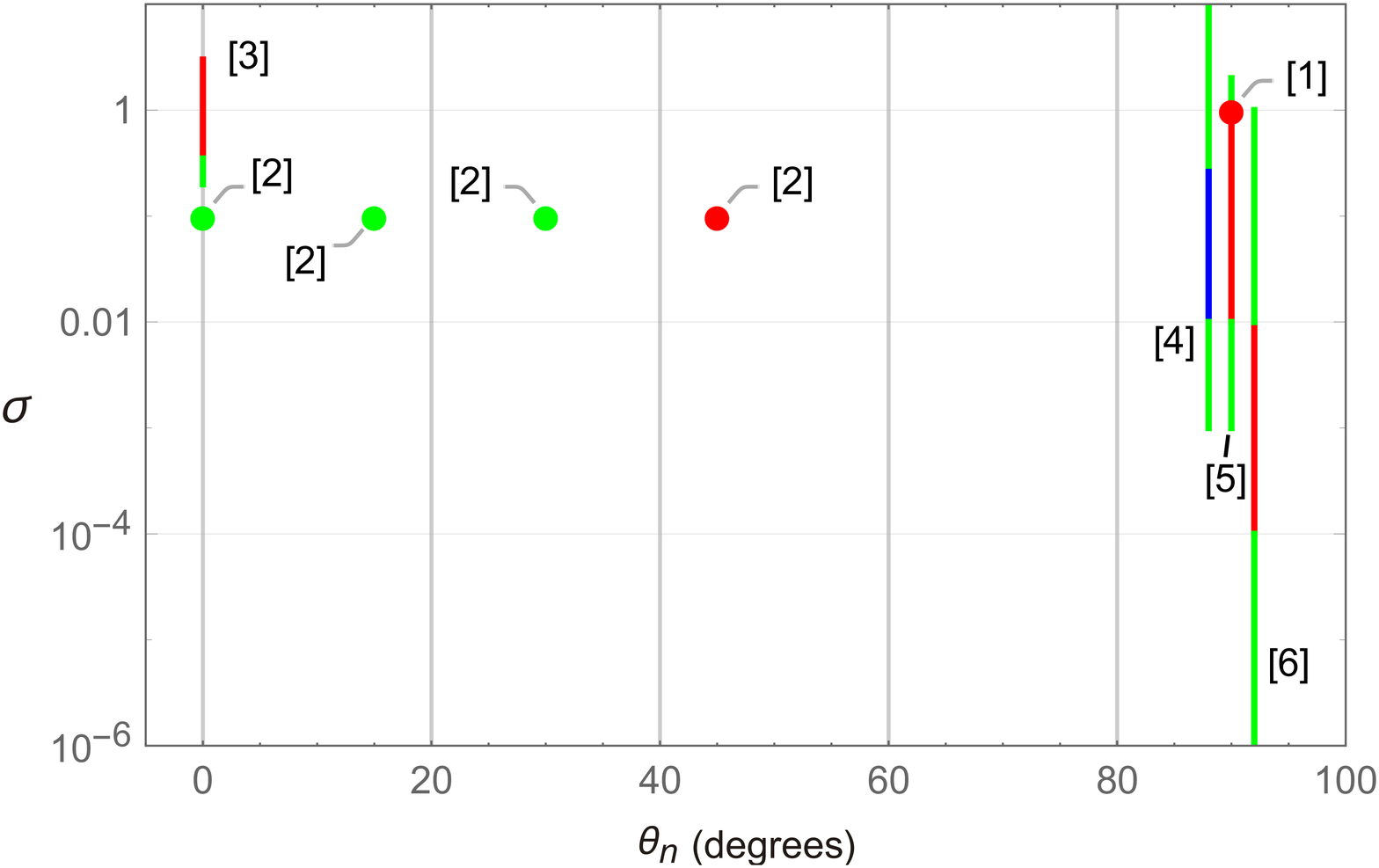}
\end{center}
\caption{Values of $\Delta_{RH}$ defined by Eq. (\ref{eq:delta0}) for the 6 articles dealing with magnetized shocks in pair plasmas, in terms of $\theta_n$ and $\sigma$. Red  means mean negative departure $\Delta_{RH} <-4\%$. Blue means \emph{positive} departure. Green  means the shock fits the MHD density jump to within $4\%$. The labels correspond to the references: [1] = \cite{2005AIPC..801..345S}, [2] = \cite{2009ApJ...698.1523S}, [3] = \cite{BretJPP2017}, [4] = \cite{1992ApJ...391...73G}, [5] = \cite{2017ApJ...840...52I}, [6] = \cite{2018MNRAS.477.5238P}. Parametric studies are indicated by vertical lines. [4,5,6] are all for $\theta_n=90^\circ$.}\label{fig:magne_pairs}
\end{figure}

\subsection{Magnetized shocks}
In a magnetized media, we have one more source of departure from MHD. There, collisionless shocks can still accelerate particles which will break the ``everything upstream goes downstream'' MHD assumption. But now in addition, the field can sustain stable pressure anisotropies and prompt departures from isotropic MHD. Considering the shock behaviour strongly depends on the external field strength and on its orientation, the physics of  magnetized collisionless shocks is extremely rich.

The upstream field strength is measured by the $\sigma$ parameter defined by Eq. (\ref{eq:sigma}). The field orientation is measured by the angle $\theta_n$ it makes with the shock front normal. In the non-relativistic regime, this angle is Lorentz invariant in the direction of the shock propagation up to order $(v/c)^2$, where $v$ is the speed of the frame to which the field is transformed. In the relativistic regime, a perpendicular or a parallel field remain so in any frame. The only articles mentioned here where an oblique field is considered in a relativistic setting are \cite{2009ApJ...698.1523S,2011ApJ...726...75S}. There, the angle of the upstream field with the shock normal is given in the simulation frame, that is, the downstream frame.

\begin{center}
\begin{table}[t]
  \caption{Simulation results for Refs. [1,4,5,6] of Figure \ref{fig:magne_pairs}. $\Delta_{RH,max}$ is the maximum relative deviation from the RH density jump (see Eq. \ref{eq:delta0}). This deviation is reached for $\sigma=\sigma_m$. $\gamma_1$ is the upstream Lorentz factor measured in the downstream frame.}
  \begin{tabular}{ l | c | c | c | c }
  \hline
Ref.               &  [1]      &  [4]            &  [5]     &  [6]         \\
  \hline
$\Delta_{RH,max}$  &  $-50\%$  &  $+27\%$        &  $-9\%$  &  $-4\%$      \\
$\sigma_m$         &  $>0.1$   &  $0.1$         &  $0.3$  &  $2.10^{-3}$ \\
$\gamma_1$         &  15       &  40 \& $10^6$  &  40      &  10          \\
\hline
  \end{tabular}\label{table}
\end{table}
\end{center}

\subsubsection{Magnetized shocks in pair plasmas}\label{sec:pairsB0}
Figure \ref{fig:magne_pairs} summarizes the results for the 6 articles falling into the present category in terms of $\theta_n$ and $\sigma$ (see references in Figure \ref{fig:magne_pairs} caption).

In [2] \citep{2009ApJ...698.1523S} the departure is about $-3\%$ for $\theta_n \leq 30^\circ$ and goes down to $-13\%$ for $45^\circ$.

In [3] \citep{BretJPP2017}, the  field is parallel and introduces a downstream anisotropy responsible for the departure from the MHD jump. For $\sigma=3$, the jump was reduced by $-35\%$, an effect all the more interesting than for a parallel shock, the MHD jump does not depend on the field\footnote{See \cite{Lichnerowicz1976} or \cite{Kulsrud2005}, Chapter 6, Eq. (36) with $B_y=0$.}.

The departure in [1] \citep{2005AIPC..801..345S} is directly related to the perpendicular field, and the density jump is said to saturate at 2 instead of 4  for larger $\sigma$'s.

The perpendicular shocks of Refs. [4,5,6] are  intriguing. The $\sigma$-ranges of MHD departure of [4,5] and [6] do not overlap. [4] \citep{1992ApJ...391...73G} finds an \emph{increase} of the density jump reaching a maximum of $+27\%$ for $\sigma = 0.1$. [5] \citep{2017ApJ...840...52I}  finds a decrease of the density jump reaching $ -9\%$ for $\sigma = 0.3$. Finally, [6] \citep{2018MNRAS.477.5238P} also finds a decrease reaching $-4\%$ for $\sigma = 2.10^{-3}$.

Table \ref{table} summarises the main features of these works. We shall comment further on these results in Section \ref{sec:iso}.

\begin{figure*}
\begin{center}
 \includegraphics[width=0.95\textwidth]{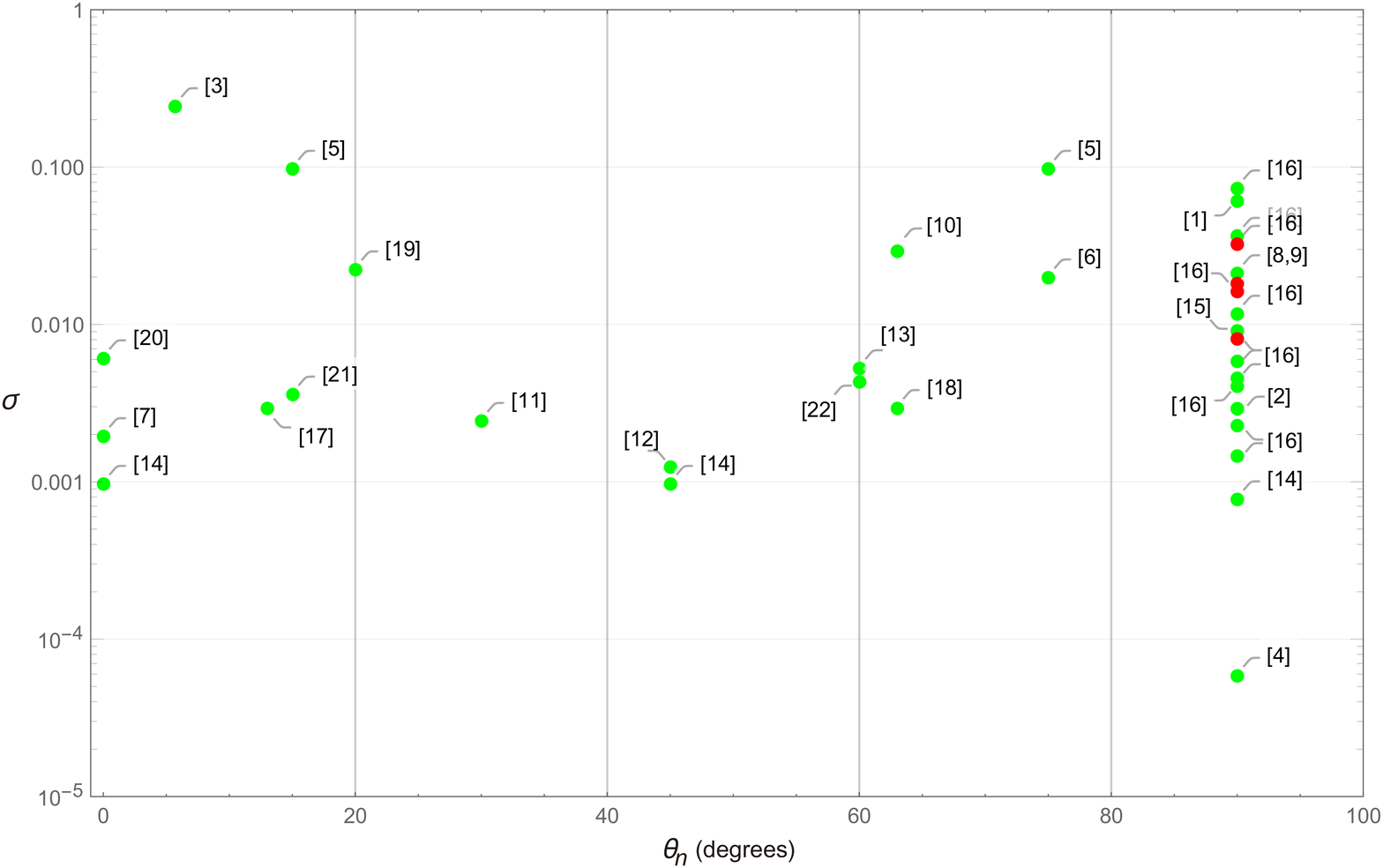}
\end{center}
\caption{Values of $\Delta_{RH}$ defined by Eq. (\ref{eq:delta0}) for the 22 articles dealing with magnetized shocks in electron/ion plasmas, in terms of $\theta_n$ and $\sigma$. The green circle means the shock fits MHD jump condition. The red circle means it does not (departure $<-5\%$). The labels correspond to the references: [1] = \cite{2003JGRA..108.1182N}, [2] = \cite{2006ApJ...652.1297L}, [3] = \cite{2010A&A...509A..89D}, [4] = \cite{2010ApJ...721..828K}, [5] = \cite{2011ApJ...726...75S}, [5] = \cite{2011ApJ...726...75S}, [6] = \cite{2011ApJ...733...63R}, [7] = \cite{2012ApJ...759...73N}, [8] = \cite{2012PhPl...19f2904P}, [9] = \cite{2014ApJ...794..153G}, [10] = \cite{2014ApJ...797...47G}, [11] = \cite{2015PhRvL.114h5003P}, [12] = \cite{2016ApJ...820...62W}, [13] = \cite{2017ApJ...846..113N}, [14] = \cite{2017ApJ...847...71B}, [15] = \cite{2017ApJ...851..134G}, [16] = \cite{2018ApJ...858...95G}, [17] = \cite{2018ApJ...864..105H}, [18] = \cite{2019ApJ...876...79K}, [19] = \cite{2019HEDP...3300709O}, [20] = \cite{2019PhPl...26c2106Z}, [21] = \cite{2019RAA....19..182F}, [22] = \cite{2020arXiv200104945L}.}
\label{fig:magne_ei}
\end{figure*}

\subsubsection{Magnetized shocks in electron/ion plasmas}
Here 22 articles were analyzed, from parallel to normal orientations and $\sigma$'s ranging from $6.10^{-5}$ \citep{2010ApJ...721..828K} to 0.25  \citep{2010A&A...509A..89D}.

Figure \ref{fig:magne_ei} pictures the results in the $(\theta_n,\sigma)$ phase space. Besides some PIC simulations performed in  \cite{2018ApJ...858...95G}, all the simulations fulfilled the RH density jump. As specified earlier, a comparison with RH was not the point of some works so that a few percent discrepancy may have escaped the analysis.

\cite{2018ApJ...858...95G} did perform a detailed comparison with the RH jump for 16 simulations\footnote{\cite{2018ApJ...858...95G} does not measure the field in terms of $\sigma$ but in terms of $\beta_{p0}=16\pi n_0k_BT_0/B_0^2$. For the purpose of the present study, we compute the $\sigma$ used in \cite{2018ApJ...858...95G} from the formula for the Alfv\'{e}nic Mach number $M_A$ given below Eq. (4) of \cite{2018ApJ...858...95G}, $M_A=M_s\sqrt{\Gamma\beta_{p0}/2}$. We then take $\sigma=1/M_A^2$.}. Discrepancies with RH range from -0.6\% (run ``Ms5beta8'') to -7\% (run ``Ms3beta8''). Only discrepancies $<-5\%$ have been highlighted in red in Figure \ref{fig:magne_ei}. The dispersion observed for some identical values of $\sigma$ stems from different values of the upstream parameters $\beta_{p0}=16\pi n_0k_BT_0/B_0^2$ (see Section \ref{sec:iso}).

In summary all the RH-departure in the examined articles come from the field and decrease the density jump\footnote{See Section \ref{sec:iso} and \cite{2018MNRAS.477.5238P} for a discussion of the jump increase in \cite{1992ApJ...391...73G}}. Jump increase stemming from accelerated particles seem to demand a few $10^3 \Omega_{ci}^{-1}$ (see Section \ref{sec:hybrid}) to be observed while the longest simulation in the present section was ran up to $559 \Omega_{ci}^{-1}$ \citep{2019RAA....19..182F} where $\Omega_{ci}$ is the ionic cyclotron frequency.

  \begin{figure}
\begin{center}
 \includegraphics[width=0.45\textwidth]{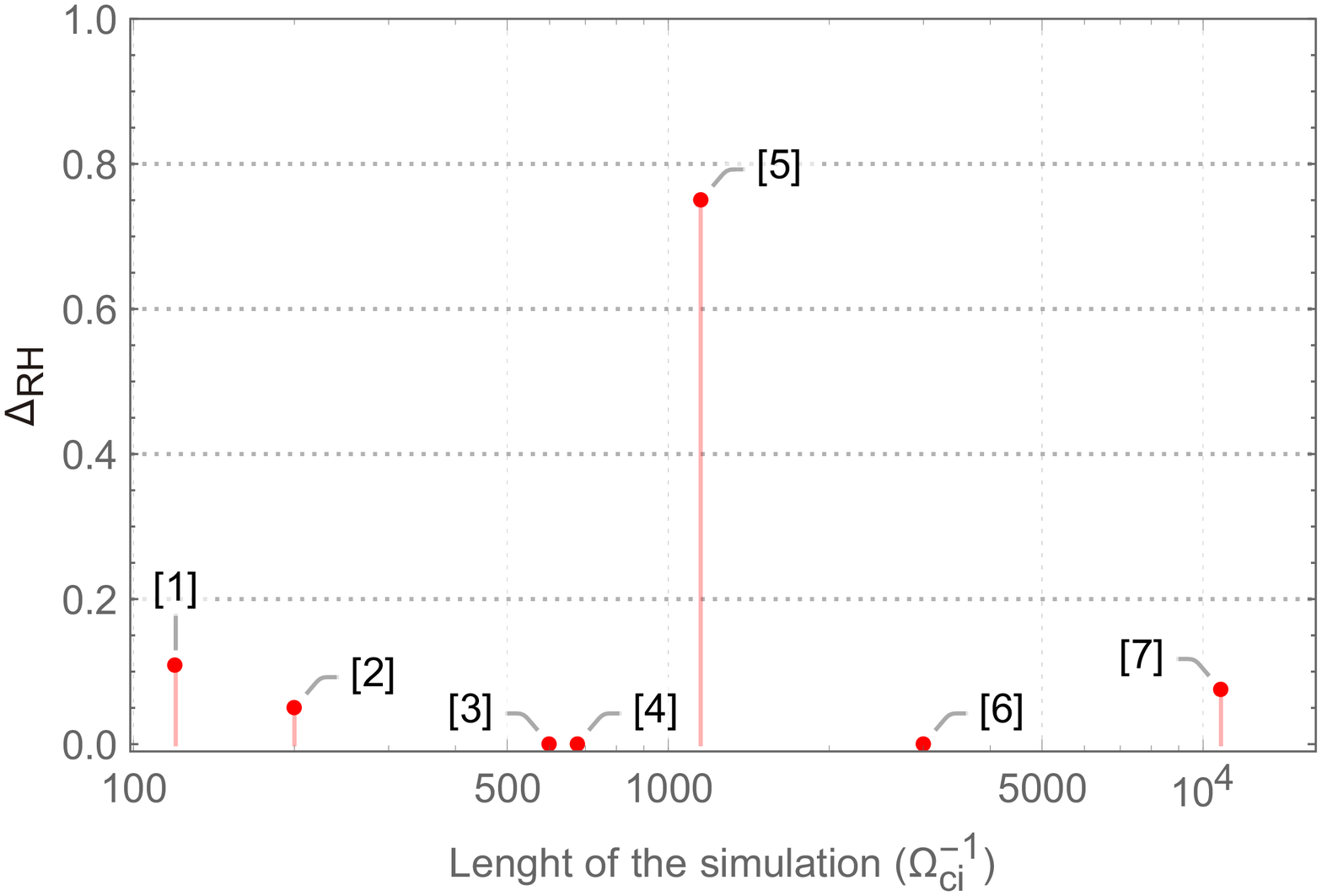}
\end{center}
\caption{Values of $\Delta_{RH}$ defined by Eq. (\ref{eq:delta0})  in terms of the simulated time lengths for various Hybrid references. The labels correspond to the references: [1] = \cite{2013ApJ...773..158G}, [2] = \cite{2014ApJ...783...91C}, [3] = \cite{2012ApJ...744...67G}, [4] = \cite{2018PPCF...60a4017C}, [5] = \cite{2019ICRC...36..209C}, [6] = \cite{2011PhPl...18b2302S}, [7] = \cite{2015ApJ...809...55B}.}
\label{fig:hybrid}
\end{figure}

\subsection{Hybrid results}\label{sec:hybrid}
Hybrid codes treat part of the medium  as a fluid, and the rest through the PIC method. In some, the fluid part is the plasma while the PIC part is devoted to accelerated particles \citep{2015ApJ...809...55B, 2018PPCF...60a4017C, 2018MNRAS.473.3394V}. In others, the electrons are the fluid part while the ions are dealt with with PIC \citep{2011PhPl...18b2302S, 2012ApJ...744...67G, 2013ApJ...773..158G, 2014ApJ...783...91C, 2019ApJ...887..165H, 2019ICRC...36..209C}.

The advantage of the method is clearly that it allows to run the simulations longer at a similar computational cost. Such a feature is necessary to render the back-reaction of accelerated particles on the shock itself. Indeed, among the articles examined, the only ones who ran the simulations longer than $10^3$ ion cyclotron periods $\Omega_{ci}^{-1}$ are Hybrid, with the 2 longest simulations, \cite{2015ApJ...809...55B} and \cite{2019ApJ...887..165H}, pushing the computation up to $10800 \Omega_{ci}^{-1}$ and $6000 \Omega_{ci}^{-1}$ respectively.

\cite{2019ApJ...887..165H} and \cite{2014ApJ...783...91C} indicate a downstream Maxwellian reaching only 80\% of the expected MHD temperature after $6000$ and $2500 \Omega_{ci}^{-1}$ respectively, due to ``energy leakage'' into accelerated particles. Regarding the density jump, Figure \ref{fig:hybrid} shows its increase  in terms of the simulated time lengths for various works. A significant difference (from +0 to +75\%) is noticeable between \cite{2019ApJ...887..165H}\footnote{The density jump in \cite{2019ApJ...887..165H} can  be inferred approximately from its Figure 3 and seems to fit RH. However, it must be somewhat higher, as the downstream temperature is lower than its RH value. Due to this uncertainty on the measured density jump, this reference is not listed on the present Figure \ref{fig:hybrid}.} and \cite{2019ICRC...36..209C}, due to the way the fluid electrons are modeled. Such is a challenge of Hybrid simulations: giving up a first principles (PIC) description of the electrons is the price to pay for longer simulation times. Much depends then of the fluid closure implemented, as evidenced by these two references.

\section{Modelling of the density jump}\label{sec:theo}

Deviations from the RH density jump observed so far are small (except maybe \cite{2019ICRC...36..209C}, see Section \ref{sec:hybrid}). We can therefore devise a first order modelling of $\Delta_{RH}(\sigma, \alpha)$, the departure from RH given by Eq. (\ref{eq:delta0}),   writing
\begin{equation}\label{eq:delta}
 \Delta_{RH}\left(\sigma, \alpha \right) \sim  \alpha \frac{\partial\Delta_{RH}}{\partial\alpha} + \sigma \frac{\partial\Delta_{RH}}{\partial\sigma},
\end{equation}
where all derivatives are considered in $(\sigma, \alpha)=(0,0)$.

 The $\alpha$ parameter determines the departure from RH due to accelerated particles. From Eq. (\ref{eq:alpha}), we see it is proportional to $F_E$, the escaping energy flux. In turn, it is known that the ability of shocks to accelerate particles depends on their magnetization, the angle of the field with the shock normal or the sonic Mach number $\mathcal{M}_1$. In addition, $F_E$ is also an increasing function of time. Therefore, strictly speaking, we should write $\alpha=\alpha(\sigma, \theta_n,\mathcal{M}_1,t)$.  A theory of the density jump accounting for all these parameters is out of the scope of this work. Yet, among them the time variable is prominent since,
\begin{equation}
\alpha(\sigma, \theta_n,\mathcal{M}_1,t=0)=0, ~~ \forall (\sigma, \theta_n,\mathcal{M}_1).
\end{equation}
She shall therefore focus on the time dependence  of $\alpha$, thus deriving its order of magnitude instead of its precise value. We will elaborate further on the effects of accelerate particles in Section \ref{sec:accel}.

\subsection{Field effect on the density jump}\label{sec:iso}
To which extent can we assume isotropic distribution functions in a collisionless plasma? Here it seems relevant to single out the magnetized and un-magnetized cases.

\emph{In a un-magnetized plasma}, an anisotropic distribution function is Weibel unstable \citep{Weibel,Kalman1968}. Although Weibel's result was only obtained for Maxwellian distribution functions with anisotropic temperatures, it seems reasonable to conjecture that any anisotropic distribution function is unstable (see \cite{SilvaAPS2019,SilvaAPS2019paper} for an effort toward a mathematical proof). We could also refer to the ample literature on collisionless  ``Maxwellianization''  (see \cite{BretJPP2015} and references therein), starting with the ``Langmuir paradox'' \citep{Langmuir1925}.

Indeed, observations of the solar wind show that in the small field limit (high $\beta_\parallel$), the protons temperature becomes isotropic \citep{BalePRL2009,SchlickeiserPRL2011,MarucaPRL2011}. It seems therefore than past the front turbulence, the downstream anisotropy of a collisionless shock should relax to isotropy on a time scale related to the instability growth rate.

\emph{The magnetized case} is different because a magnetized Vlasov plasma can sustain stable anisotropies \citep{Gary1993}. A strong enough field $\mathbf{B}_0$ can therefore maintain an anisotropic upstream and/or an anisotropic downstream. An isotropic upstream can turn anisotropic as it goes to the downstream depending on the magnetization parameter $\sigma$.

How should a residual downstream anisotropy modify the density jump? A hint can be given by the fact that the field tends to reduce the degrees of freedom $D$ of the plasma. It therefore increases its macroscopic adiabatic index $\Gamma = 1+2/D$. And since the RH density jump is a decreasing function of $\Gamma$, the presence of the field should lower it. This has been seen in the literature review.

A quantitative assessment of this reduction implies determining the downstream anisotropy in terms of the upstream properties. As noted earlier, the effect is especially interesting for the parallel case because the MHD jump of a parallel shock is $\sigma$-independent. Making an ansatz on the kinetic evolution of the plasma through the front, \cite{BretJPP2018} derived for a parallel shock in a non-relativistic pair plasma (strong shock limit, upstream $\Gamma=5/3$)\footnote{See $r_+$ of Eq. (3.5) of \cite{BretJPP2018}, with $\chi_1=\infty$.},
\begin{eqnarray}\label{eq:jump_sigma_para}
  r &=& \frac{1}{2} \left(-\sigma+\sqrt{(\sigma-9) (\sigma-1)}+5\right)  \nonumber \\
  &=& 4 - \frac{4}{3}\sigma + \mathcal{O}(\sigma^2).
\end{eqnarray}
After some algebra we get in Eq. (\ref{eq:delta})
\begin{equation}\label{eq:delta_para}
\frac{\partial\Delta_{RH}}{\partial\sigma} = -\frac{1}{3}.
\end{equation}

The perpendicular case is quite different as the MHD jump is already reduced by the field like\footnote{Taylor expansion of Eq. (7) of \cite{BretPoP2019}, with $\sigma=M_{A1}^{-2}$.},
\begin{equation}\label{eq:MHDjump_sigma_perp}
  r = 4 - 18\sigma + \mathcal{O}(\sigma^2).
\end{equation}
Applying the same method than for the parallel case, \citet{BretPoP2019} derived for the perpendicular one  (see Appendix \ref{ap1}),
\begin{equation}\label{eq:Jump_pop19}
  r = 4 - \frac{86}{3}\sigma + \mathcal{O}(\sigma^2),
\end{equation}
where $86/3\sim 28.6$ represents therefore a steeper decline than the MHD one (\ref{eq:MHDjump_sigma_perp}). Here, we can write in Eq. (\ref{eq:delta}) after some algebra,
\begin{equation}\label{eq:delta_perp}
\frac{\partial\Delta_{RH}}{\partial\sigma} = -\frac{8}{3}.
\end{equation}

At the present stage it is premature to accurately contrast the model with the simulations, since a full blown theory should account for the composition of the plasmas (pair or e/i) and relativistic effects. Yet, the orders of magnitude can be checked at least with \cite{2018ApJ...858...95G} and \cite{BretJPP2017}. The latter is relativist while the former is not.

  \begin{figure}
\begin{center}
 \includegraphics[width=0.45\textwidth]{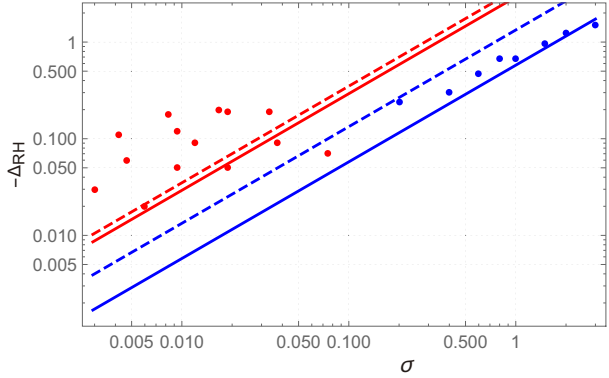}
\end{center}
\caption{Values of $-\Delta_{RH}$ defined by Eq. (\ref{eq:delta0})  in terms of $\sigma$, for \cite{2018ApJ...858...95G} (red) and \cite{BretJPP2017} (blue). Dashed lines stem from Eqs. (\ref{eq:delta_para},\ref{eq:delta_perp}). Plain lines = best fits of the form $\Delta_{RH}=a\sigma$, with $a\in \mathbb{R}$.}
\label{fig:guo_bret}
\end{figure}

The values\footnote{Computed from the data recorded in Table 4 of \cite{2018ApJ...858...95G}.} of $\Delta_{RH}$ obtained in \cite{2018ApJ...858...95G} for a perpendicular shock can be fitted by $\Delta_{RH} \sim -1.14\sigma$.

The values\footnote{Computed from Figure 4a of \cite{BretJPP2017}.} of $\Delta_{RH}$ obtained in \cite{BretJPP2017} for a parallel shock can be fitted by $\Delta_{RH} \sim -0.13\sigma$.
The orders of magnitude fit well Eqs. (\ref{eq:delta_para}) for the parallel case and (\ref{eq:delta_perp}) for the perpendicular one.

These results are displayed on Figure \ref{fig:guo_bret}. The dispersion around the fit for \cite{2018ApJ...858...95G} (red) is important due to the various values of $\beta_{p0}$ explored. The dispersion around the fit for \cite{BretJPP2017} is far less important because  all the runs dealt with plasmas initially cold.

We may finally comment on the ``departure windows'' observed on Figure \ref{fig:magne_pairs} for pair perpendicular shocks, and on the $+27\%$ \emph{increase} of the jump observed in \cite{1992ApJ...391...73G} (ref [4] of  the present Figure \ref{fig:magne_pairs}). This increase was attributed to the emission of electromagnetic waves at the shock front. As commented in \cite{2018MNRAS.477.5238P} (ref [6] of the present Figure \ref{fig:magne_pairs}), this should be a dimension effect as \cite{1992ApJ...391...73G} is the only 1D simulation of the 4 references.

The ``departure windows''\footnote{The ``window`'' in \cite{2017ApJ...840...52I} in visible on Figure 15 of \cite{2017ApJ...840...52I}.} in [5,6] \citep{2017ApJ...840...52I,2018MNRAS.477.5238P} were left unexplained. Besides their weak amplitude (-9 and -4\% respectively\footnote{Determined from Figure 2 of \cite{2017ApJ...840...52I} and from Figure 2 of \cite{2018MNRAS.477.5238P}.} for the maximum departure) they may simply arise from the following process: at small $\sigma$ the jump is in agreement with RH, as evidenced in all the simulations and discussed in Section \ref{sec:iso}. Then the field generates an anisotropy which triggers a negative departure from the isotropic MHD jump. Yet, for even higher $\sigma$'s, the growing anisotropy drives the jump to 2, and MHD does the very same (see Figure 6 in \cite{1992ApJ...391...73G} and Figure 2 in \cite{2018MNRAS.477.5238P}). We may therefore expect departure windows for perpendicular shocks since MHD and PIC simulations have the same limits in both the weak and the strong field limits.

  \begin{figure}
\begin{center}
 \includegraphics[width=0.45\textwidth]{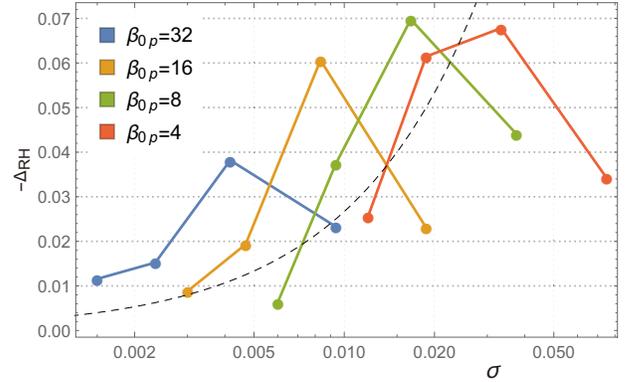}
\end{center}
\caption{Values of $-\Delta_{RH}$ defined by Eq. (\ref{eq:delta0})  in terms of $\sigma$ for the runs listed in Table 4 of \cite{2018ApJ...858...95G}. The dashed line stems from Eq. (\ref{eq:delta_perp}). For a given $\beta_{p0}$, an MHD-departure window is clear.}
\label{fig:guo_window}
\end{figure}

This is confirmed on Figure \ref{fig:guo_window} where we plotted again Figure \ref{fig:guo_bret} for \cite{2018ApJ...858...95G}, simply joining the points sharing the same $\beta_{p0}$. An MHD-departure \emph{window} is evidenced for each curves, centered around values of $\sigma$  similar to those observed on Figure \ref{fig:magne_pairs} for pair plasmas.

Such a feature is absent in parallel shocks since MHD is insensitive to the field. In parallel geometry, we can just expect a departure \emph{threshold} beyond which kinetic effects progressively drive the collisionless density jump away from its MHD counterpart.

\subsection{Accelerated particles effect on the density jump}\label{sec:accel}
In a fluid shock particles constantly share their energy with each others through binary collisions. In collisionless shocks, it has been known for long that particles can be accelerated \citep{Krymskii1977,Axford1977,Blandford78,Bell1978a,Bell1978b}. By going back and forth around the front and/or escaping upstream, these particles escape the Rankine-Hugoniot budget. \cite{BretJPP2018a,BretLPB2018} derived some requirements for particle accelerations, that are all fulfilled in the present cases.

A simple calculation derived from \cite{1999ApJ...526..385B} allows to conclude that accelerated particles should increase the density jump. We outline it here for completeness. Considering the non-relativistic regime for simplicity, we start writing the conservation equations between the upstream and the downstream with subscripts ``1'' and ``2'' respectively. Labeling $n_i,v_i,P_i,\Gamma$ the density, velocity pressure and adiabatic index of the fluid we have,
\begin{eqnarray}
                                             n_2v_2 &=& n_1v_1 - F_m,    \label{eq:RH_CR1}\\
                                       n_2v_2^2+P_2 &=& n_1v_1^2+P_1 - F_p ,\label{eq:RH_CR2}\\
  \frac{1}{2}n_2v_2^3+\frac{\Gamma}{\Gamma-1}P_2v_2 &=& \frac{1}{2}n_1v_1^3+\frac{\Gamma}{\Gamma-1}P_1v_1 - F_E, \label{eq:RH_CR3}
\end{eqnarray}
where $F_m,F_p,F_E$ are the mass, momentum and energy fluxes escaping the Rankine-Hugoniot budget because of accelerated particles. It turns out that $n_1v_1 \gg F_m$, $n_1v_1^2 \gg  F_p$ while $F_E$ in Eq. (\ref{eq:RH_CR3}) is \emph{not} negligible with respect to $n_1v_1^3$ (see \cite{1999ApJ...526..385B} and references therein). We can therefore neglect $F_m,F_p$ in Eqs. (\ref{eq:RH_CR1},\ref{eq:RH_CR2}) and solve the system for $n_2$. From Eq. (\ref{eq:RH_CR1}) we derive $v_2=(n_1/n_2)v_1$. Using this expression to eliminate $v_2$ from Eqs. (\ref{eq:RH_CR2},\ref{eq:RH_CR3}) allows to derive two different expressions for $P_2$. Equaling them yields a second degree polynomial in $n_2$ that can be solved exactly. The shock solution reads,
\begin{equation}\label{eq:jump_CR}
  r = \frac{1+\Gamma  \mathcal{M}_1^2+\sqrt{\mathcal{M}_1^4 \left(\alpha  \left(\Gamma^2-1\right)+1\right)-2 \mathcal{M}_1^2+1}}{(1-\alpha) (\Gamma -1) \mathcal{M}_1^2+2},
\end{equation}
where $\alpha$ is defined by Eq. (\ref{eq:alpha}) and,
\begin{equation}\label{eq:Machnb}
\mathcal{M}_1^2=\frac{n_1v_1^2}{\Gamma P_1},
\end{equation}
 is the upstream sonic Mach number. Clearly the density jump (\ref{eq:jump_CR}) can be arbitrarily high as $\alpha\rightarrow 1$. Such a feature can be elaborated further by modeling $\alpha$ \citep{1999ApJ...526..385B,2010ApJ...722.1727V}.

In the strong shock limit $\mathcal{M}_1\rightarrow\infty$ it reduces to,
\begin{eqnarray}\label{eq:jump_CR_Strong}
  r_\infty &=& \frac{\sqrt{\alpha  (\Gamma^2-1)+1}+\Gamma }{(1-\alpha ) (\Gamma -1)}, \nonumber \\
                        &= & \frac{\Gamma +1}{\Gamma -1} + \alpha\frac{  \Gamma ^2+2 \Gamma +1}{2 (\Gamma -1)} + \mathcal{O}(\alpha^2)\nonumber \\
                        &= & 4  + \frac{16}{3}\alpha + \mathcal{O}(\alpha^2)~~\text{for}~~\Gamma=5/3.
\end{eqnarray}
After soma algebra we find in Eq. (\ref{eq:delta}),
\begin{equation}
\frac{\partial\Delta_{RH}}{\partial\alpha} = +\frac{4}{3},
\end{equation}
so that the relative deviation stemming from accelerated particles should read $+\frac{4}{3}\alpha$. Now, $\alpha$ is not constant in time because the energy $F_E$ poured into cosmic rays is not. For example, the maximum energy of accelerated particles grows like $t^{1/2}$ for relativistic shocks \citep{Sironi2013,2018MNRAS.477.5238P} and $t$ for non-relativistic ones \citep{2014ApJ...794...47C}.

These results allow to phenomenologically assess the $\alpha$ coefficient in (\ref{eq:jump_CR_Strong}). Density jump departures stemming from accelerated particles were notified in \cite{2009ApJ...693L.127K} and \cite{2012PPCF...54l5004S} (see Section \ref{sec:littnoB0})\footnote{ \cite{2012ApJ...759...73N} ran their (un-magnetized) simulation up to $4111\omega_{pi}^{-1}$ and saw no sign of particle acceleration. Yet, the authors themselves found it odd as they wrote in the conclusion: ``In PIC simulations one uses few computational particles to represent very many real electrons or ions and thus introduces artificial collisionality. Would that impact, and possibly prevent, particle pre-acceleration in our simulations?''.}. For magnetized shocks, such departures were notified in the references featured in Figure \ref{fig:hybrid}, among which \cite{2019ICRC...36..209C} stands out as the only one where the density jump is clearly evaluated at various times.   These results suggest altogether that the departure reaches a few percents for run times of the order of $5.10^3$ time units. For magnetized shocks, this ``time unit'' is the gyro-period. For the un-magnetized shocks in pair plasmas, it is the inverse electronic plasma frequency.

The orders of magnitude derived from un-magnetized shocks in pairs \citep{2009ApJ...693L.127K, 2012PPCF...54l5004S} and magnetized parallel shocks in electron/ion \citep{2019ICRC...36..209C}, are similar. To which extent can they be generalized to any shock? Indeed, these 3 studies are but a sample of the possible combinations achievable varying the parameters $(\sigma, \theta_n,\mathcal{M}_1)$ in pairs and electron/ion. It turns out that the acceleration efficiency has been studied extensively in magnetized pair \citep{2009ApJ...698.1523S} or electron/ion plasmas \citep{2011ApJ...726...75S, 2014ApJ...783...91C, 2014ApJ...794..153G, 2018ApJ...858...95G}, and it was found that the aforementioned order of magnitude is representative of the full range of possibilities \citep{SironiPrivate}.

Assuming that the departure grows like $\beta t^\kappa$ with $\kappa > 0$, we then set,
\begin{equation}\label{eq:kappa}
\frac{4}{3}\beta t^\kappa = \mathcal{O}(10^{-1})~~\mathrm{for}~~t = \mathcal{O}(5.10^3) ,
\end{equation}
so that,
\begin{eqnarray}
\alpha &=& \mathcal{O}(5^{-\kappa}10^{-1-3\kappa})t^\kappa , \nonumber \\
       &=& \mathcal{O}(10^{-1-3.7\kappa})t^\kappa,
\end{eqnarray}
where $t$ is measured in the dominant unit of the simulation.   Note that the scaling of the maximum energy of the accelerated particles does not have to translate to the energy flux $F_E$ (see Eq. \ref{eq:alpha}) leaking into accelerated particles, since the most energetic ones are but a few. In all the papers examined, the only one from which it was possible to extract a time dependent variation of $F_E$ is  \cite{2019ICRC...36..209C}. Its Figure 2 suggests $\kappa = \mathcal{O}(1)$. Further works would be welcome to narrow down the value of $\kappa$ and the time scale of $5.10^3$ set by  Eq. (\ref{eq:kappa}).

\subsection{Summary}
Gathering the results obtained in Section \ref{sec:iso} for the field effects, and in Section \ref{sec:accel} for accelerated particles, we can complete Eq. (\ref{eq:delta}) and write,
\begin{equation}\label{eq:deltaOK1}
 \Delta_{RH}(\sigma, \alpha) \sim  +\mathcal{O}(10^{-1-3.7\kappa})t - \sigma \left\{\begin{array}{l}
                                                                             1/3~~\theta_n=0, \\
                                                                             8/3~~\theta_n=\pi/2.
                                                                           \end{array}
   \right.
\end{equation}

Since we are interested in the order of magnitude of the field correction, we can aggregate the results for $\theta_n=0$ and $\pi/2$  and propose,
\begin{equation}\label{eq:deltaOK}
 \Delta_{RH}(\sigma, \alpha) \sim  +  \mathcal{O}(10^{-1-3.7\kappa})t^\kappa - \sigma \mathcal{O}(1),
\end{equation}
  where $\kappa$ is of order unity. As commented above, the first term, $+\mathcal{O}(10^{-1-3.7\kappa})t^\kappa$, should be representative, \emph{in order of magnitude}, of the full spectrum of shocks populating the $(\sigma, \theta_n,\mathcal{M}_1)$ phase space.

 To which extent can we make the same claim for the correction term $\propto \sigma$, due to the field driven anisotropy?

The field correction obviously vanishes for $\sigma=0$, $\forall ( \theta_n,\mathcal{M}_1)$. Regarding the $\theta_n$ variation, Figures \ref{fig:magne_pairs} and \ref{fig:magne_ei} present the results of 9 simulations at various angles in pair plasmas, and  31 in electron/ion plasmas. No difference of order of magnitude has been detected with respect to Eq. (\ref{eq:deltaOK}) for the coefficient of $\sigma$.

As for the incidence of $\mathcal{M}_1$, all the pair shocks featuring Figure \ref{fig:magne_pairs} are strong, that is, $\mathcal{M}_1 \gg 1$. As for those in electron/ion presented on Figure \ref{fig:magne_ei}, they span sonic Mach numbers ranging from 2 \citep{2018ApJ...858...95G, 2018ApJ...864..105H, 2019ApJ...876...79K} to $\mathcal{M}_1 \gg 1$. Again no deviation, order of magnitude wise, has been detected from the coefficient of the $\sigma$ correction reported in Eq. (\ref{eq:deltaOK}).

The only part of the phase space parameter which has not been tested in the literature is the deviation from RH in weak shocks in pair plasmas. If the model developed in \cite{BretJPP2018,BretPoP2019,bretLPB2020} is further confirmed by PIC simulation, then this gap of weak shock in pairs will be filled.

\section{Summary and conclusion}

The present work represents an attempt to determine when the RH density jump can be applied to collisionless shocks. A tentative answer is given by Eq. (\ref{eq:deltaOK}) which is valid as long as $\Delta_{RH}$, the relative departure from the RH density jump, is small.

The departure is the sum of 2 terms. One, positive, arises from the accelerated particles escaping the RH budget, and grows with time. The second is negative and stems from the pressure anisotropy sustained by the field.

The literature review clearly evidences positive departures arising from particles acceleration, and negative ones from field driven anisotropies. We didn't find studies considering both effects together, contemplating for example the possibility that they compensate each other.

What about the long times and/or strong field regime, beyond the validity of Eq. (\ref{eq:deltaOK})?

Can accelerated particles drive a density jump arbitrary high on the long run? Some theoretical models suggests so \citep{1999ApJ...526..385B,2010ApJ...722.1727V} together with some simulations \citep{2019ICRC...36..209C}. Yet, since the maximum energy of these particles saturates with time \citep{2013ApJ...771...54S}, the energy flux $F_E$  escaping the Rankine-Hugoniot budget in Eq. (\ref{eq:RH_CR3}) could also saturate.

As for the large $\sigma$ regime, we have to distinguish parallel shocks from perpendicular ones.

For parallel shocks, the MHD jump is insensitive to the field. The jump departure is therefore going to increase until it saturates since a large $\sigma$ will eventually trigger an anisotropy yielding an asymptotic density jump. For example, for a strong shock in 3D with $\Gamma=5/3$, the MHD jump remains 4 while the collisionless one tends to 2, resulting in the largest possible departure $\Delta_{RH}= -50\%$.

For perpendicular shocks, the large $\sigma$ jumps are identical for MHD and collisionless shocks. The negative jump departure should therefore reach a maximum for intermediate values of $\sigma$ and vanish before and after. Such a ``departure window'' has been retrieved for relativistic pair shocks (see \cite{2017ApJ...840...52I,2018MNRAS.477.5238P} and Figure \ref{fig:magne_pairs} of the present work) and non-relativistic electron/ion shocks \citep{2018ApJ...858...95G}. In the later case, Figure \ref{fig:guo_window} shows that the location and magnitude of the window depend on the upstream proton temperature parameter $\beta_{p0}$. At any rate, $\Delta_{RH}$ never goes below $-10\%$ in any of the aforementioned studies.

The dimensionality of the works involved in the present study may be its main limitation. Even though the formalism of \cite{BretJPP2018,BretPoP2019,bretLPB2020} is 3D,  simulations under scrutiny feature at best 3 velocity dimensions but only 2 spatial dimensions (2D3V). To with extent can the conclusions be generalized to 3D space? The reduced number of spatial dimensions can have important effects on particle acceleration and/or the external field effect. Comparisons between 2D and 3D results found for example that some particles trapping occur in 2D and not in 3D \citep{cruz2017,Trotta2018}. Also, considering only 2 spatial dimensions necessary excludes waves and instabilities with a wave vector $\mathbf{k}$ oriented along the excluded spatial dimension. Indeed, theoretical explorations of the full unstable $\mathbf{k}$-spectrum of some beam-plasma (or Weibel-like) instabilities, found it is truly 3D \citep{Kalman1968,2008NJPh...10a3029D,BretPoP2014a,Stockem2016}.

Finally, \cite{MatsumotoPRL2017} (injection scheme, Mach number $=22.8$) explicitly compared electron acceleration in 2D and 3D simulations for quasi-perpendicular electron/ion shocks. They found the acceleration to be more efficient in 3D than in 2D, be it with an out-of-plane or an  in-plane field\footnote{Figure 4d of \cite{MatsumotoPRL2017} suggests that the total amount of energy in accelerated electrons could be closer to the 3D case for 2D in-plane, than for 2D out-of-plane simulations.}.

A very similar system was studied in  \cite{2014ApJ...794..153G} (reflecting wall, Mach number $=3$). There, 2D simulations were also compared to 3D ones. Yet, in contrast with \cite{MatsumotoPRL2017}, it was found that the 2D in-plane field configuration ``is a good choice to capture the acceleration physics of the full 3D problem''. Perhaps the discrepancy with  \cite{MatsumotoPRL2017} is due to the different Mach numbers or the different methods. Therefore, it seems that as concluded in \cite{2017ApJ...847...71B}, ``true 3D simulations are urgently needed to resolve this issue''.

\acknowledgments

This work has been  achieved under projects  ENE2016-75703-R from the Spanish Ministerio de Econom\'{\i}a y Competitividad and SBPLY/17/180501/000264 from the Junta de Comunidades de Castilla-La Mancha. Thanks are due to Colby  Haggerty, Thales Silva, Arno Vanthieghem, Laurent Gremillet, Damiano Caprioli and Lorenzo Sironi for valuable inputs.

\appendix
\section{    Proof of Eq. (10)  } \label{ap1}
We start from Eq. (25) of \cite{BretPoP2019} where the parameter $\chi_1$ is proportional to the sonic Mach number. This equation already has the adiabatic index $\Gamma=5/3$. The equation for the density jump $r$ in the strong shock limit $\chi_1\rightarrow\infty$ is given by the coefficient of $\chi_1^2$ of Eq. (25) in \cite{BretPoP2019}. The result is the 3rd degree polynomial,
\begin{equation}
P(r) \equiv (2 r (r+2)-5) r \sigma +2 (r-5) r+8 = 0,
\end{equation}
where it can be checked that for $\sigma=0$, the 2 roots are $r=1$ and 4. We then set $r=4+k\sigma$ with $k\in \mathbb{R}$, and perform a Taylor expansion of $P(4-k\sigma)$ up to first order in $\sigma$. The zeroth order vanishes, and the first order also if $k=-86/3$, proving Eq. (\ref{eq:Jump_pop19}).

\bibliography{BibBret}{}

\begin{thebibliography}{}
\expandafter\ifx\csname natexlab\endcsname\relax\def\natexlab#1{#1}\fi
\providecommand{\url}[1]{\href{#1}{#1}}
\providecommand{\dodoi}[1]{doi:~\href{http://doi.org/#1}{\nolinkurl{#1}}}
\providecommand{\doeprint}[1]{\href{http://ascl.net/#1}{\nolinkurl{http://ascl.net/#1}}}
\providecommand{\doarXiv}[1]{\href{https://arxiv.org/abs/#1}{\nolinkurl{https://arxiv.org/abs/#1}}}

\bibitem[{Ardaneh {et~al.}(2016)Ardaneh, Cai, \& Nishikawa}]{Ardaneh2016}
Ardaneh, K., Cai, D., \& Nishikawa, K.-I. 2016, The Astrophysical Journal, 827,
  124, \dodoi{10.3847/0004-637x/827/2/124}

\bibitem[{Axford {et~al.}(1977)Axford, Leer, \& Skadron}]{Axford1977}
Axford, W., Leer, E., \& Skadron, G. 1977, Proc. 15th International Cosmic Ray
  Conference, 11, 132

\bibitem[{{Bai} {et~al.}(2015){Bai}, {Caprioli}, {Sironi}, \&
  {Spitkovsky}}]{2015ApJ...809...55B}
{Bai}, X.-N., {Caprioli}, D., {Sironi}, L., \& {Spitkovsky}, A. 2015, \apj,
  809, 55, \dodoi{10.1088/0004-637X/809/1/55}

\bibitem[{Bale {et~al.}(2009)Bale, Kasper, Howes, Quataert, Salem, \&
  Sundkvist}]{BalePRL2009}
Bale, S.~D., Kasper, J.~C., Howes, G.~G., {et~al.} 2009, Phys. Rev. Lett., 103,
  211101, \dodoi{10.1103/PhysRevLett.103.211101}

\bibitem[{Bale {et~al.}(2003)Bale, Mozer, \& Horbury}]{PRLBow1}
Bale, S.~D., Mozer, F.~S., \& Horbury, T.~S. 2003, Phys. Rev. Lett., 91, 265004

\bibitem[{Balogh \& Treumann(2013)}]{balogh2013}
Balogh, A., \& Treumann, R. 2013, Physics of Collisionless Shocks: Space Plasma
  Shock Waves, ISSI Scientific Report Series (Springer New York).
\newblock \url{https://books.google.es/books?id=mR4\_AAAAQBAJ}

\bibitem[{{Bell}(1978{\natexlab{a}})}]{Bell1978a}
{Bell}, A.~R. 1978{\natexlab{a}}, Mon. Not. R. Astron. Soc, 182, 147

\bibitem[{{Bell}(1978{\natexlab{b}})}]{Bell1978b}
---. 1978{\natexlab{b}}, Mon. Not. R. Astron. Soc, 182, 443

\bibitem[{{Berezhko} \& {Ellison}(1999)}]{1999ApJ...526..385B}
{Berezhko}, E.~G., \& {Ellison}, D.~C. 1999, \apj, 526, 385,
  \dodoi{10.1086/307993}

\bibitem[{Blandford \& Ostriker(1978)}]{Blandford78}
Blandford, R., \& Ostriker, J. 1978, Astrophysical Journal, 221, L29

\bibitem[{{Bohdan} {et~al.}(2017){Bohdan}, {Niemiec}, {Kobzar}, \&
  {Pohl}}]{2017ApJ...847...71B}
{Bohdan}, A., {Niemiec}, J., {Kobzar}, O., \& {Pohl}, M. 2017, \apj, 847, 71,
  \dodoi{10.3847/1538-4357/aa872a}

\bibitem[{Bret(2014)}]{BretPoP2014a}
Bret, A. 2014, Physics of Plasmas, 21, 022106

\bibitem[{{Bret}(2015)}]{BretJPP2015}
{Bret}, A. 2015, Journal of Plasma Physics, 81, 455810202

\bibitem[{{Bret} \& {Narayan}(2018)}]{BretJPP2018}
{Bret}, A., \& {Narayan}, R. 2018, Journal of Plasma Physics, 84, 905840604,
  \dodoi{10.1017/S0022377818001125}

\bibitem[{{Bret} \& {Narayan}(2019)}]{BretPoP2019}
---. 2019, Physics of Plasmas, 26, 062108, \dodoi{10.1063/1.5099000}

\bibitem[{Bret \& Narayan(2020)}]{bretLPB2020}
Bret, A., \& Narayan, R. 2020, Laser and Particle Beams, 1–7,
  \dodoi{10.1017/S0263034620000117}

\bibitem[{{Bret} \& {Pe'er}(2018{\natexlab{a}})}]{BretJPP2018a}
{Bret}, A., \& {Pe'er}, A. 2018{\natexlab{a}}, Journal of Plasma Physics, 84,
  905840311, \dodoi{10.1017/S0022377818000636}

\bibitem[{{Bret} \& {Pe'er}(2018{\natexlab{b}})}]{BretLPB2018}
---. 2018{\natexlab{b}}, Laser and Particle Beams, 36, 458,
  \dodoi{10.1017/S0263034618000472}

\bibitem[{{Bret} {et~al.}(2017){Bret}, {Pe'er}, {Sironi}, {S\c{a}dowski}, \&
  {Narayan}}]{BretJPP2017}
{Bret}, A., {Pe'er}, A., {Sironi}, L., {S\c{a}dowski}, A., \& {Narayan}, R.
  2017, Journal of Plasma Physics, 83, 715830201,
  \dodoi{10.1017/S0022377817000290}

\bibitem[{{Bret} {et~al.}(2013){Bret}, {Stockem}, {Fiuza}, {Ruyer},
  {Gremillet}, {Narayan}, \& {Silva}}]{2013PhPl...20d2102B}
{Bret}, A., {Stockem}, A., {Fiuza}, F., {et~al.} 2013, Physics of Plasmas, 20,
  042102, \dodoi{10.1063/1.4798541}

\bibitem[{{Bret} {et~al.}(2014){Bret}, {Stockem}, {Narayan}, \&
  {Silva}}]{2014PhPl...21g2301B}
{Bret}, A., {Stockem}, A., {Narayan}, R., \& {Silva}, L.~O. 2014, Physics of
  Plasmas, 21, 072301, \dodoi{10.1063/1.4886121}

\bibitem[{{Caprioli} \& {Haggerty}(2019)}]{2019ICRC...36..209C}
{Caprioli}, D., \& {Haggerty}, C. 2019, in International Cosmic Ray Conference,
  Vol.~36, 36th International Cosmic Ray Conference (ICRC2019), 209.
\newblock \doarXiv{1909.06288}

\bibitem[{{Caprioli} \& {Spitkovsky}(2014{\natexlab{a}})}]{2014ApJ...783...91C}
{Caprioli}, D., \& {Spitkovsky}, A. 2014{\natexlab{a}}, \apj, 783, 91,
  \dodoi{10.1088/0004-637X/783/2/91}

\bibitem[{{Caprioli} \& {Spitkovsky}(2014{\natexlab{b}})}]{2014ApJ...794...47C}
---. 2014{\natexlab{b}}, \apj, 794, 47, \dodoi{10.1088/0004-637X/794/1/47}

\bibitem[{{Casse} {et~al.}(2018){Casse}, {van Marle}, \&
  {Marcowith}}]{2018PPCF...60a4017C}
{Casse}, F., {van Marle}, A.~J., \& {Marcowith}, A. 2018, Plasma Physics and
  Controlled Fusion, 60, 014017, \dodoi{10.1088/1361-6587/aa8482}

\bibitem[{{Chang} {et~al.}(2008){Chang}, {Spitkovsky}, \&
  {Arons}}]{2008ApJ...674..378C}
{Chang}, P., {Spitkovsky}, A., \& {Arons}, J. 2008, \apj, 674, 378,
  \dodoi{10.1086/524764}

\bibitem[{Cruz {et~al.}(2017)Cruz, Alves, Bamford, Bingham, Fonseca, \&
  Silva}]{cruz2017}
Cruz, F., Alves, E.~P., Bamford, R.~A., {et~al.} 2017, Physics of Plasmas, 24,
  022901, \dodoi{10.1063/1.4975310}

\bibitem[{{Dieckmann} \& {Bret}(2017)}]{2017JPlPh..83a9004D}
{Dieckmann}, M.~E., \& {Bret}, A. 2017, Journal of Plasma Physics, 83,
  905830104, \dodoi{10.1017/S0022377816001288}

\bibitem[{{Dieckmann} \& {Bret}(2018)}]{2018MNRAS.473..198D}
---. 2018, \mnras, 473, 198, \dodoi{10.1093/mnras/stx2387}

\bibitem[{{Dieckmann} {et~al.}(2008){Dieckmann}, {Bret}, \&
  {Shukla}}]{2008NJPh...10a3029D}
{Dieckmann}, M.~E., {Bret}, A., \& {Shukla}, P.~K. 2008, New Journal of
  Physics, 10, 013029, \dodoi{10.1088/1367-2630/10/1/013029}

\bibitem[{{Dieckmann} {et~al.}(2019){Dieckmann}, {Folini}, {Hotz}, {Nordman},
  {Dell'Acqua}, {Ynnerman}, \& {Walder}}]{DieckmannAA2019}
{Dieckmann}, M.~E., {Folini}, D., {Hotz}, I., {et~al.} 2019, \aap, 621, A142,
  \dodoi{10.1051/0004-6361/201834393}

\bibitem[{{Dieckmann} {et~al.}(2010{\natexlab{a}}){Dieckmann}, {Murphy},
  {Meli}, \& {Drury}}]{2010A&A...509A..89D}
{Dieckmann}, M.~E., {Murphy}, G.~C., {Meli}, A., \& {Drury}, L.~O.~C.
  2010{\natexlab{a}}, \aap, 509, A89, \dodoi{10.1051/0004-6361/200912643}

\bibitem[{{Dieckmann} {et~al.}(2016){Dieckmann}, {Sarri}, {Doria}, {Ynnerman},
  \& {Borghesi}}]{2016PhPl...23f2111D}
{Dieckmann}, M.~E., {Sarri}, G., {Doria}, D., {Ynnerman}, A., \& {Borghesi}, M.
  2016, Physics of Plasmas, 23, 062111, \dodoi{10.1063/1.4953568}

\bibitem[{{Dieckmann} {et~al.}(2010{\natexlab{b}}){Dieckmann}, {Sarri},
  {Romagnani}, {Kourakis}, \& {Borghesi}}]{2010PPCF...52b5001D}
{Dieckmann}, M.~E., {Sarri}, G., {Romagnani}, L., {Kourakis}, I., \&
  {Borghesi}, M. 2010{\natexlab{b}}, Plasma Physics and Controlled Fusion, 52,
  025001, \dodoi{10.1088/0741-3335/52/2/025001}

\bibitem[{{Erkaev} {et~al.}(2000){Erkaev}, {Vogl}, \& {Biernat}}]{Erkaev2000}
{Erkaev}, N.~V., {Vogl}, D.~F., \& {Biernat}, H.~K. 2000, Journal of Plasma
  Physics, 64, 561, \dodoi{10.1017/S002237780000893X}

\bibitem[{{Fang} {et~al.}(2019){Fang}, {Lu}, {Yan}, \&
  {Yu}}]{2019RAA....19..182F}
{Fang}, J., {Lu}, C.-Y., {Yan}, J.-W., \& {Yu}, H. 2019, Research in Astronomy
  and Astrophysics, 19, 182, \dodoi{10.1088/1674-4527/19/12/182}

\bibitem[{Fitzpatrick(2014)}]{fitzpatrick2014plasma}
Fitzpatrick, R. 2014, Plasma Physics: An Introduction (Taylor \& Francis).
\newblock \url{https://books.google.es/books?id=0RwbBAAAQBAJ}

\bibitem[{{Fiuza} {et~al.}(2012){Fiuza}, {Fonseca}, {Tonge}, {Mori}, \&
  {Silva}}]{2012PhRvL.108w5004F}
{Fiuza}, F., {Fonseca}, R.~A., {Tonge}, J., {Mori}, W.~B., \& {Silva}, L.~O.
  2012, \prl, 108, 235004, \dodoi{10.1103/PhysRevLett.108.235004}

\bibitem[{{Gallant} {et~al.}(1992){Gallant}, {Hoshino}, {Langdon}, {Arons}, \&
  {Max}}]{1992ApJ...391...73G}
{Gallant}, Y.~A., {Hoshino}, M., {Langdon}, A.~B., {Arons}, J., \& {Max}, C.~E.
  1992, \apj, 391, 73, \dodoi{10.1086/171326}

\bibitem[{{Gargat{\'e}} \& {Spitkovsky}(2012)}]{2012ApJ...744...67G}
{Gargat{\'e}}, L., \& {Spitkovsky}, A. 2012, \apj, 744, 67,
  \dodoi{10.1088/0004-637X/744/1/67}

\bibitem[{Gary(1993)}]{Gary1993}
Gary, S. 1993, Theory of Space Plasma Microinstabilities, Cambridge Atmospheric
  and Space Science Series (Cambridge University Press)

\bibitem[{Gerbig \& Schlickeiser(2011)}]{Gerbig2011}
Gerbig, D., \& Schlickeiser, R. 2011, The Astrophysical Journal, 733, 32.
\newblock \url{http://stacks.iop.org/0004-637X/733/i=1/a=32}

\bibitem[{Goedbloed {et~al.}(2019)Goedbloed, Keppens, \& Poedts}]{keppens2019}
Goedbloed, H., Keppens, R., \& Poedts, S. 2019, Magnetohydrodynamics of
  Laboratory and Astrophysical Plasmas (Cambridge University Press),
  \dodoi{10.1017/9781316403679}

\bibitem[{{Guo} \& {Giacalone}(2013)}]{2013ApJ...773..158G}
{Guo}, F., \& {Giacalone}, J. 2013, \apj, 773, 158,
  \dodoi{10.1088/0004-637X/773/2/158}

\bibitem[{{Guo} {et~al.}(2014{\natexlab{a}}){Guo}, {Sironi}, \&
  {Narayan}}]{2014ApJ...794..153G}
{Guo}, X., {Sironi}, L., \& {Narayan}, R. 2014{\natexlab{a}}, \apj, 794, 153,
  \dodoi{10.1088/0004-637X/794/2/153}

\bibitem[{{Guo} {et~al.}(2014{\natexlab{b}}){Guo}, {Sironi}, \&
  {Narayan}}]{2014ApJ...797...47G}
---. 2014{\natexlab{b}}, \apj, 797, 47, \dodoi{10.1088/0004-637X/797/1/47}

\bibitem[{{Guo} {et~al.}(2017){Guo}, {Sironi}, \&
  {Narayan}}]{2017ApJ...851..134G}
---. 2017, \apj, 851, 134, \dodoi{10.3847/1538-4357/aa9b82}

\bibitem[{{Guo} {et~al.}(2018){Guo}, {Sironi}, \&
  {Narayan}}]{2018ApJ...858...95G}
---. 2018, \apj, 858, 95, \dodoi{10.3847/1538-4357/aab6ad}

\bibitem[{{Ha} {et~al.}(2018){Ha}, {Ryu}, {Kang}, \& {van
  Marle}}]{2018ApJ...864..105H}
{Ha}, J.-H., {Ryu}, D., {Kang}, H., \& {van Marle}, A.~J. 2018, \apj, 864, 105,
  \dodoi{10.3847/1538-4357/aad634}

\bibitem[{{Haggerty} \& {Caprioli}(2019)}]{2019ApJ...887..165H}
{Haggerty}, C.~C., \& {Caprioli}, D. 2019, \apj, 887, 165,
  \dodoi{10.3847/1538-4357/ab58c8}

\bibitem[{{Iwamoto} {et~al.}(2017){Iwamoto}, {Amano}, {Hoshino}, \&
  {Matsumoto}}]{2017ApJ...840...52I}
{Iwamoto}, M., {Amano}, T., {Hoshino}, M., \& {Matsumoto}, Y. 2017, \apj, 840,
  52, \dodoi{10.3847/1538-4357/aa6d6f}

\bibitem[{Johnson \& Cheret(1998)}]{johnson1998classic}
Johnson, J., \& Cheret, R. 1998, Classic Papers in Shock Compression Science:
  Edition en anglais, High pressure shock compression of condensed matter
  (Springer).
\newblock \url{https://books.google.es/books?id=256ws-XBjzwC}

\bibitem[{Kalman {et~al.}(1968)Kalman, Montes, \& Quemada}]{Kalman1968}
Kalman, B.~G., Montes, C., \& Quemada, D. 1968, Phys. Fluids, 11, 1797

\bibitem[{{Kang} {et~al.}(2019){Kang}, {Ryu}, \& {Ha}}]{2019ApJ...876...79K}
{Kang}, H., {Ryu}, D., \& {Ha}, J.-H. 2019, \apj, 876, 79,
  \dodoi{10.3847/1538-4357/ab16d1}

\bibitem[{Karimabadi {et~al.}(1995)Karimabadi, Krauss-Varban, \&
  Omidi}]{Karimabadi95}
Karimabadi, H., Krauss-Varban, D., \& Omidi, N. 1995, Geophysical Research
  Letters, 22, 2689, \dodoi{10.1029/95GL02788}

\bibitem[{{Kato}(2007)}]{2007ApJ...668..974K}
{Kato}, T.~N. 2007, \apj, 668, 974, \dodoi{10.1086/521297}

\bibitem[{{Kato} \& {Takabe}(2008)}]{2008ApJ...681L..93K}
{Kato}, T.~N., \& {Takabe}, H. 2008, \apjl, 681, L93, \dodoi{10.1086/590387}

\bibitem[{{Kato} \& {Takabe}(2010)}]{2010ApJ...721..828K}
---. 2010, \apj, 721, 828, \dodoi{10.1088/0004-637X/721/1/828}

\bibitem[{{Keshet} {et~al.}(2009){Keshet}, {Katz}, {Spitkovsky}, \&
  {Waxman}}]{2009ApJ...693L.127K}
{Keshet}, U., {Katz}, B., {Spitkovsky}, A., \& {Waxman}, E. 2009, \apjl, 693,
  L127, \dodoi{10.1088/0004-637X/693/2/L127}

\bibitem[{Krymskii(1977)}]{Krymskii1977}
Krymskii, G. 1977, Doklady Akademii Nauk SSSR, 234, 1306

\bibitem[{Kulsrud(2005)}]{Kulsrud2005}
Kulsrud, R.~M. 2005, Plasma Physics for Astrophysics (Princeton, NJ: Princeton
  Univ. Press)

\bibitem[{Langmuir(1925)}]{Langmuir1925}
Langmuir, I. 1925, Phys. Rev., 26, 585

\bibitem[{{Lemoine} {et~al.}(2019{\natexlab{a}}){Lemoine}, {Gremillet},
  {Pelletier}, \& {Vanthieghem}}]{2019PhRvL.123c5101L}
{Lemoine}, M., {Gremillet}, L., {Pelletier}, G., \& {Vanthieghem}, A.
  2019{\natexlab{a}}, \prl, 123, 035101, \dodoi{10.1103/PhysRevLett.123.035101}

\bibitem[{{Lemoine} {et~al.}(2019{\natexlab{b}}){Lemoine}, {Pelletier},
  {Vanthieghem}, \& {Gremillet}}]{2019PhRvE.100c3210L}
{Lemoine}, M., {Pelletier}, G., {Vanthieghem}, A., \& {Gremillet}, L.
  2019{\natexlab{b}}, \pre, 100, 033210, \dodoi{10.1103/PhysRevE.100.033210}

\bibitem[{{Lemoine} {et~al.}(2019{\natexlab{c}}){Lemoine}, {Vanthieghem},
  {Pelletier}, \& {Gremillet}}]{2019PhRvE.100c3209L}
{Lemoine}, M., {Vanthieghem}, A., {Pelletier}, G., \& {Gremillet}, L.
  2019{\natexlab{c}}, \pre, 100, 033209, \dodoi{10.1103/PhysRevE.100.033209}

\bibitem[{{Lezhnin} {et~al.}(2020){Lezhnin}, {Fox}, {Schaeffer}, {Matteucci},
  {Bhattacharjee}, {Spitkovsky}, \& {Germaschewski}}]{2020arXiv200104945L}
{Lezhnin}, K.~V., {Fox}, W., {Schaeffer}, D.~B., {et~al.} 2020, arXiv e-prints,
  arXiv:2001.04945.
\newblock \doarXiv{2001.04945}

\bibitem[{{Li} {et~al.}(2017){Li}, {Zhou}, {Huang}, {Qiao}, \&
  {He}}]{2017PhPl...24d2113L}
{Li}, R., {Zhou}, C.~T., {Huang}, T.~W., {Qiao}, B., \& {He}, X.~T. 2017,
  Physics of Plasmas, 24, 042113, \dodoi{10.1063/1.4980832}

\bibitem[{{Lichnerowicz}(1976)}]{Lichnerowicz1976}
{Lichnerowicz}, A. 1976, Journal of Mathematical Physics, 17, 2135,
  \dodoi{10.1063/1.522857}

\bibitem[{{Lyubarsky}(2006)}]{2006ApJ...652.1297L}
{Lyubarsky}, Y. 2006, \apj, 652, 1297, \dodoi{10.1086/508606}

\bibitem[{{Martins} {et~al.}(2009){Martins}, {Fonseca}, {Silva}, \&
  {Mori}}]{2009ApJ...695L.189M}
{Martins}, S.~F., {Fonseca}, R.~A., {Silva}, L.~O., \& {Mori}, W.~B. 2009,
  \apjl, 695, L189, \dodoi{10.1088/0004-637X/695/2/L189}

\bibitem[{Maruca {et~al.}(2011)Maruca, Kasper, \& Bale}]{MarucaPRL2011}
Maruca, B.~A., Kasper, J.~C., \& Bale, S.~D. 2011, Phys. Rev. Lett., 107,
  201101, \dodoi{10.1103/PhysRevLett.107.201101}

\bibitem[{Matsumoto {et~al.}(2017)Matsumoto, Amano, Kato, \&
  Hoshino}]{MatsumotoPRL2017}
Matsumoto, Y., Amano, T., Kato, T.~N., \& Hoshino, M. 2017, Phys. Rev. Lett.,
  119, 105101, \dodoi{10.1103/PhysRevLett.119.105101}

\bibitem[{Mihalas \& Weibel-Mihalas(1999)}]{mihalas1999}
Mihalas, D., \& Weibel-Mihalas, B. 1999, Foundations of Radiation
  Hydrodynamics, Dover Books on Physics (Dover Publications).
\newblock \url{https://books.google.es/books?id=C3h3DQAAQBAJ}

\bibitem[{{Moreno} {et~al.}(2020){Moreno}, {Dieckmann}, {Folini}, {Walder},
  {Ribeyre}, {Tikhonchuk}, \& {d'Humi{\`e}res}}]{2020PPCF...62b5022M}
{Moreno}, Q., {Dieckmann}, M.~E., {Folini}, D., {et~al.} 2020, Plasma Physics
  and Controlled Fusion, 62, 025022, \dodoi{10.1088/1361-6587/ab5bfb}

\bibitem[{{Nakanotani} {et~al.}(2017){Nakanotani}, {Matsukiyo}, {Hada}, \&
  {Mazelle}}]{2017ApJ...846..113N}
{Nakanotani}, M., {Matsukiyo}, S., {Hada}, T., \& {Mazelle}, C.~X. 2017, \apj,
  846, 113, \dodoi{10.3847/1538-4357/aa8363}

\bibitem[{{Naseri} {et~al.}(2018){Naseri}, {Bochkarev}, {Ruan}, {Bychenkov},
  {Khudik}, \& {Shvets}}]{2018PhPl...25a2118N}
{Naseri}, N., {Bochkarev}, S.~G., {Ruan}, P., {et~al.} 2018, Physics of
  Plasmas, 25, 012118, \dodoi{10.1063/1.5008278}

\bibitem[{{Niemiec} {et~al.}(2012){Niemiec}, {Pohl}, {Bret}, \& {Wieland
  }}]{2012ApJ...759...73N}
{Niemiec}, J., {Pohl}, M., {Bret}, A., \& {Wieland }, V. 2012, \apj, 759, 73,
  \dodoi{10.1088/0004-637X/759/1/73}

\bibitem[{{Nishikawa} {et~al.}(2009){Nishikawa}, {Niemiec}, {Hardee},
  {Medvedev}, {Sol}, {Mizuno}, {Zhang}, {Pohl}, {Oka}, \&
  {Hartmann}}]{NishikawaApJL2009}
{Nishikawa}, K.~I., {Niemiec}, J., {Hardee}, P.~E., {et~al.} 2009, \apjl, 698,
  L10, \dodoi{10.1088/0004-637X/698/1/L10}

\bibitem[{{Nishimura} {et~al.}(2003){Nishimura}, {Matsumoto}, {Kojima}, \&
  {Gary}}]{2003JGRA..108.1182N}
{Nishimura}, K., {Matsumoto}, H., {Kojima}, H., \& {Gary}, S.~P. 2003, Journal
  of Geophysical Research (Space Physics), 108, 1182,
  \dodoi{10.1029/2002JA009671}

\bibitem[{Novo {et~al.}(2016)Novo, Bret, \& Sinha}]{Stockem2016}
Novo, A.~S., Bret, A., \& Sinha, U. 2016, New Journal of Physics, 18, 105002

\bibitem[{{Otsuka} {et~al.}(2019){Otsuka}, {Matsukiyo}, \&
  {Hada}}]{2019HEDP...3300709O}
{Otsuka}, F., {Matsukiyo}, S., \& {Hada}, T. 2019, High Energy Density Physics,
  33, 100709, \dodoi{10.1016/j.hedp.2019.100709}

\bibitem[{{Park} {et~al.}(2015){Park}, {Caprioli}, \&
  {Spitkovsky}}]{2015PhRvL.114h5003P}
{Park}, J., {Caprioli}, D., \& {Spitkovsky}, A. 2015, \prl, 114, 085003,
  \dodoi{10.1103/PhysRevLett.114.085003}

\bibitem[{{Park} {et~al.}(2012){Park}, {Workman}, {Blackman}, {Ren}, \&
  {Siller}}]{2012PhPl...19f2904P}
{Park}, J., {Workman}, J.~C., {Blackman}, E.~G., {Ren}, C., \& {Siller}, R.
  2012, Physics of Plasmas, 19, 062904, \dodoi{10.1063/1.4729913}

\bibitem[{{Pelletier} {et~al.}(2019){Pelletier}, {Gremillet}, {Vanthieghem}, \&
  {Lemoine}}]{2019PhRvE.100a3205P}
{Pelletier}, G., {Gremillet}, L., {Vanthieghem}, A., \& {Lemoine}, M. 2019,
  \pre, 100, 013205, \dodoi{10.1103/PhysRevE.100.013205}

\bibitem[{{Plotnikov} {et~al.}(2018){Plotnikov}, {Grassi}, \&
  {Grech}}]{2018MNRAS.477.5238P}
{Plotnikov}, I., {Grassi}, A., \& {Grech}, M. 2018, \mnras, 477, 5238,
  \dodoi{10.1093/mnras/sty979}

\bibitem[{{Riquelme} \& {Spitkovsky}(2011)}]{2011ApJ...733...63R}
{Riquelme}, M.~A., \& {Spitkovsky}, A. 2011, \apj, 733, 63,
  \dodoi{10.1088/0004-637X/733/1/63}

\bibitem[{{Ruyer} {et~al.}(2015){Ruyer}, {Gremillet}, \&
  {Bonnaud}}]{2015PhPl...22h2107R}
{Ruyer}, C., {Gremillet}, L., \& {Bonnaud}, G. 2015, Physics of Plasmas, 22,
  082107, \dodoi{10.1063/1.4928096}

\bibitem[{{Ruyer} {et~al.}(2017){Ruyer}, {Gremillet}, {Bonnaud}, \&
  {Riconda}}]{2017PhPl...24d1409R}
{Ruyer}, C., {Gremillet}, L., {Bonnaud}, G., \& {Riconda}, C. 2017, Physics of
  Plasmas, 24, 041409, \dodoi{10.1063/1.4979187}

\bibitem[{{Sagdeev}(1966)}]{Sagdeev66}
{Sagdeev}, R.~Z. 1966, Reviews of Plasma Physics, 4, 23

\bibitem[{Salas(2007)}]{Salas2007}
Salas, M.~D. 2007, Shock Waves, 16, 477

\bibitem[{Schlickeiser {et~al.}(2011)Schlickeiser, Michno, Ibscher, Lazar, \&
  Skoda}]{SchlickeiserPRL2011}
Schlickeiser, R., Michno, M.~J., Ibscher, D., Lazar, M., \& Skoda, T. 2011,
  Phys. Rev. Lett., 107, 201102, \dodoi{10.1103/PhysRevLett.107.201102}

\bibitem[{Schwartz {et~al.}(2011)Schwartz, Henley, Mitchell, \&
  Krasnoselskikh}]{PRLBow2}
Schwartz, S.~J., Henley, E., Mitchell, J., \& Krasnoselskikh, V. 2011, Phys.
  Rev. Lett., 107, 215002

\bibitem[{{Silva}(2020)}]{SilvaAPS2019paper}
{Silva}, T. 2020, In preparation

\bibitem[{{Silva} {et~al.}(2019){Silva}, {Afeyan}, \& {Silva}}]{SilvaAPS2019}
{Silva}, T., {Afeyan}, B., \& {Silva}, L. 2019, in APS Meeting Abstracts, Vol.
  2019, APS Division of Plasma Physics Meeting Abstracts, BO7.003

\bibitem[{Sironi(2020)}]{SironiPrivate}
Sironi, L. 2020, Private Communication

\bibitem[{{Sironi} \& {Spitkovsky}(2009{\natexlab{a}})}]{2009ApJ...698.1523S}
{Sironi}, L., \& {Spitkovsky}, A. 2009{\natexlab{a}}, \apj, 698, 1523,
  \dodoi{10.1088/0004-637X/698/2/1523}

\bibitem[{{Sironi} \& {Spitkovsky}(2009{\natexlab{b}})}]{2009ApJ...707L..92S}
---. 2009{\natexlab{b}}, \apjl, 707, L92, \dodoi{10.1088/0004-637X/707/1/L92}

\bibitem[{{Sironi} \& {Spitkovsky}(2011)}]{2011ApJ...726...75S}
---. 2011, \apj, 726, 75, \dodoi{10.1088/0004-637X/726/2/75}

\bibitem[{{Sironi} {et~al.}(2013{\natexlab{a}}){Sironi}, {Spitkovsky}, \&
  {Arons}}]{Sironi2013}
{Sironi}, L., {Spitkovsky}, A., \& {Arons}, J. 2013{\natexlab{a}}, The
  Astrophysical Journal, 771, 54, \dodoi{10.1088/0004-637X/771/1/54}

\bibitem[{{Sironi} {et~al.}(2013{\natexlab{b}}){Sironi}, {Spitkovsky}, \&
  {Arons}}]{2013ApJ...771...54S}
---. 2013{\natexlab{b}}, \apj, 771, 54, \dodoi{10.1088/0004-637X/771/1/54}

\bibitem[{{Spitkovsky}(2005)}]{2005AIPC..801..345S}
{Spitkovsky}, A. 2005, in American Institute of Physics Conference Series, Vol.
  801, Astrophysical Sources of High Energy Particles and Radiation, ed.
  T.~{Bulik}, B.~{Rudak}, \& G.~{Madejski}, 345--350, \dodoi{10.1063/1.2141897}

\bibitem[{{Spitkovsky}(2008{\natexlab{a}})}]{2008ApJ...682L...5S}
{Spitkovsky}, A. 2008{\natexlab{a}}, \apjl, 682, L5, \dodoi{10.1086/590248}

\bibitem[{{Spitkovsky}(2008{\natexlab{b}})}]{2008ApJ...673L..39S}
---. 2008{\natexlab{b}}, \apjl, 673, L39, \dodoi{10.1086/527374}

\bibitem[{{Stockem} {et~al.}(2014{\natexlab{a}}){Stockem}, {Fiuza}, {Bret},
  {Fonseca}, \& {Silva}}]{2014NatSR...4E3934S}
{Stockem}, A., {Fiuza}, F., {Bret}, A., {Fonseca}, R.~A., \& {Silva}, L.~O.
  2014{\natexlab{a}}, Scientific Reports, 4, 3934, \dodoi{10.1038/srep03934}

\bibitem[{{Stockem} {et~al.}(2012){Stockem}, {Fi{\'u}za}, {Fonseca}, \&
  {Silva}}]{2012PPCF...54l5004S}
{Stockem}, A., {Fi{\'u}za}, F., {Fonseca}, R.~A., \& {Silva}, L.~O. 2012,
  Plasma Physics and Controlled Fusion, 54, 125004,
  \dodoi{10.1088/0741-3335/54/12/125004}

\bibitem[{{Stockem} {et~al.}(2014{\natexlab{b}}){Stockem}, {Grismayer},
  {Fonseca}, \& {Silva}}]{2014PhRvL.113j5002S}
{Stockem}, A., {Grismayer}, T., {Fonseca}, R.~A., \& {Silva}, L.~O.
  2014{\natexlab{b}}, \prl, 113, 105002, \dodoi{10.1103/PhysRevLett.113.105002}

\bibitem[{{Stockem Novo} {et~al.}(2015){Stockem Novo}, {Bret}, {Fonseca}, \&
  {Silva}}]{2015ApJ...803L..29S}
{Stockem Novo}, A., {Bret}, A., {Fonseca}, R.~A., \& {Silva}, L.~O. 2015,
  \apjl, 803, L29, \dodoi{10.1088/2041-8205/803/2/L29}

\bibitem[{{Sugiyama}(2011)}]{2011PhPl...18b2302S}
{Sugiyama}, T. 2011, Physics of Plasmas, 18, 022302, \dodoi{10.1063/1.3552026}

\bibitem[{Tidman {et~al.}(1971)Tidman, Krall, \& of~Maryland}]{tidman1971shock}
Tidman, D., Krall, N., \& of~Maryland, U. 1971, Shock Waves in Collisionless
  Plasmas, Wiley Medical Publication (Wiley-Interscience).
\newblock \url{https://books.google.es/books?id=soguAAAAIAAJ}

\bibitem[{{Tomita} {et~al.}(2019){Tomita}, {Ohira}, \&
  {Yamazaki}}]{2019ApJ...886...54T}
{Tomita}, S., {Ohira}, Y., \& {Yamazaki}, R. 2019, \apj, 886, 54,
  \dodoi{10.3847/1538-4357/ab4a10}

\bibitem[{Trotta \& Burgess(2018)}]{Trotta2018}
Trotta, D., \& Burgess, D. 2018, Monthly Notices of the Royal Astronomical
  Society, 482, 1154, \dodoi{10.1093/mnras/sty2756}

\bibitem[{{van Marle} {et~al.}(2018){van Marle}, {Casse}, \&
  {Marcowith}}]{2018MNRAS.473.3394V}
{van Marle}, A.~J., {Casse}, F., \& {Marcowith}, A. 2018, \mnras, 473, 3394,
  \dodoi{10.1093/mnras/stx2509}

\bibitem[{{Vanthieghem} {et~al.}(2020){Vanthieghem}, {Lemoine}, {Plotnikov},
  {Grassi}, {Grech}, {Gremillet}, \& {Pelletier}}]{2020arXiv200201141V}
{Vanthieghem}, A., {Lemoine}, M., {Plotnikov}, I., {et~al.} 2020, arXiv
  e-prints, arXiv:2002.01141.
\newblock \doarXiv{2002.01141}

\bibitem[{{Vink} {et~al.}(2010){Vink}, {Yamazaki}, {Helder}, \&
  {Schure}}]{2010ApJ...722.1727V}
{Vink}, J., {Yamazaki}, R., {Helder}, E.~A., \& {Schure}, K.~M. 2010, \apj,
  722, 1727, \dodoi{10.1088/0004-637X/722/2/1727}

\bibitem[{Vogl {et~al.}(2001)Vogl, Biernat, Erkaev, Farrugia, \&
  M\"uhlbachler}]{Vogl2001}
Vogl, D.~F., Biernat, H.~K., Erkaev, N.~V., Farrugia, C.~J., \& M\"uhlbachler,
  S. 2001, Nonlinear Processes in Geophysics, 8, 167,
  \dodoi{10.5194/npg-8-167-2001}

\bibitem[{Weibel(1959)}]{Weibel}
Weibel, E.~S. 1959, Phys. Rev. Lett., 2, 83

\bibitem[{{Wieland} {et~al.}(2016){Wieland}, {Pohl}, {Niemiec}, {Rafighi}, \&
  {Nishikawa}}]{2016ApJ...820...62W}
{Wieland}, V., {Pohl}, M., {Niemiec}, J., {Rafighi}, I., \& {Nishikawa}, K.-I.
  2016, \apj, 820, 62, \dodoi{10.3847/0004-637X/820/1/62}

\bibitem[{{Zekovi{\'c}}(2019)}]{2019PhPl...26c2106Z}
{Zekovi{\'c}}, V. 2019, Physics of Plasmas, 26, 032106,
  \dodoi{10.1063/1.5050909}

\bibitem[{Zel'dovich \& Raizer(2002)}]{Zeldovich}
Zel'dovich, I., \& Raizer, Y. 2002, Physics of Shock Waves and High-Temperature
  Hydrodynamic Phenomena, Dover Books on Physics (Dover Publications)

\end{thebibliography}
\bibliographystyle{aasjournal}

\end{document}